# Ferromagnetism and Curie temperature of Vanadium-doped Nitrides

**Van An Dinh\* and Hiroshi Katayama-Yoshida**
Department of Condensed Matter Physics,
Department of Computational Nanomaterials Design,
Nanoscience and Nanotechnology Center,
The Institute of Scientific and Industrial Research,
Osaka University, Mihogaoka 8-1, Ibaraki, Osaka 567-0047, Japan
\*E-mail: divan@cmp.sanken.osaka-u.ac.jp

**Abstract** Electronic structures, exchange interaction mechanism between magnetic ions and Curie temperature of Vanadium - doped Nitrides (AlN, GaN, and InN) are studied within KKR-LSDA-CPA. It is found that the ferromagnetic super-exchange interaction mechanism is dominant at low concentrations of Vanadium, but the anti-ferromagnetic super-exchange interaction appears and reduces the stabilization of ferromagnetism at sufficiently high concentrations ($x > 0.10$), especially for Vanadium-doped AlN and Vanadium- doped GaN. The estimation of the Curie temperature within the mean field approximation shows the Curie temperature of Vanadium-doped Nitrides exceeding the room temperature with a few constituents of Vanadium.

**Keywords** first principle calculation, materials design, spintronics, dilute magnetic semiconductors, nitrides

## I. Introduction

The discovery of ferromagnetic dilute magnetic semiconductors (DMS) [1] by incorporating the transition metal (TM) ions into various host semiconductors has received much attention because of the potential applications that employ simultaneously the charge and spin of the carriers in the control of the electronic properties of materials. For practical applications, the magnetic ordering of DMS must be achieved in the temperature higher than room temperature. Hence, to synthesize the materials with Curie temperature higher than room temperature has become one of the hot topics in spintronics.

For III-V compound-based DMS, though the theoretical work has predicted the high ferromagnetic Curie temperature ($T_c$) in Mn-doped GaAs based DMS [2], up to now, the $T_c$ reported in the newest experiment of this material has been limited to 150K [3] which is still much lower than the room temperature. In order to use this material in industrial applications, one must enhance the Curie temperature in some ways such as the codoping method with Nitrogen or Carbon [4-6], the delta-doping method [7], etc. Recently, GaN as well as related compounds are treated as key advanced materials for spintronic devices. Nitrides have wide band gap energies, exhibit very high optical efficiencies, and they could be used in electronic devices such as high power and high frequency transitions, blue and ultraviolet (UV) light-emitting diodes, laser diodes, etc. Furthermore, theoretical calculations and experimental data reporting the ferromagnetic behavior in (Mn, Cr)-doped Nitrides at very high temperatures [2, 8-11] suggest a promising possibility of these materials for spintronic device applications. Katayama-Yoshida and Sato have predicted the stabilization of ferromagnetic phase for Vanadium (V) -doped GaN at various concentrations of Vanadium [2]. However, very little data on the V -doped AlN and V-doped InN has been reported, and the underlying mechanism of exchange interaction between magnetic ions in these compounds has not been well understood. Therefore, detailed investigation on the feasibility of V-doped Nitrides is of great importance from the viewpoint of both scientific knowledge and industrial application as well.

In this paper, the role of 3*d*-electrons of Vanadium and magnetism in III-Nitrides compound-based DMS ($Al_{1-x}V_xN$, $Ga_{1-x}V_xN$ and $In_{1-x}V_xN$) are investigated, based on first principle calculations, and material design of V-doped Nitrides is proposed.

The electronic structures are calculated by using the Korringa-Kohn-Rostoker (KKR) method combined with coherent potential approximation (CPA). The $T_c$ is evaluated by using the mean field approximation (MFA). The calculation is performed for various concentrations of Vanadium substituting randomly atoms at cation sites of wurtzite III-Nitride compounds.

The paper is organized as follows. The calculation method is presented in Sec. II. The calculations of density of states (DOS) for $Al_{1-x}V_xN$, $Ga_{1-x}V_xN$ and $In_{1-x}V_xN$ at 2%, 5% and 10% of Vanadium concentration and the estimation of Curie temperatures for various concentrations of Vanadium within the mean field approximation are given, and the exchange interaction mechanism in these compounds is also discussed in Sec. III. Finally, some concluding remarks are given in Sec. IV.

## II. Calculation scheme

The electronic structures of $Al_{1-x}V_xN$, $Ga_{1-x}V_xN$ and $In_{1-x}V_xN$ are calculated by the KKR-LDSA-CPA method developed by Akai and Dederichs in treating transition metal alloys [12] and InAs-based DMS [13]. The DMSs' are fabricated by substituting randomly the atoms at cation sites of wurtzite III-Nitride compounds by transition metal Vanadium. Self-consistent electronic structures are performed within the local spin density approximation (LSDA) with the parameterization suggested by Moruzzi *et al.* [15]. The relativistic effect is taken into account by employing the scalar relativistic approximation. The potential is treated as muffin-tin one, and the wave functions in the muffin-tin spheres are expanded up to $l=2$ in the real harmonics, where $l$ is the angular momentum defined at each site. *425 k*-points in the irreducible part of the first Brillouin zone are used in our calculations. For convenience, the lattice constants are fixed at the values of pure compounds of AlN, GaN, and InN, respectively, as shown in the Tab. I.

**TABLE I.** Lattice constants and band-gap energies of wurtzite AlN, GaN and InN taken from Ref. 14

| Compound | AlN | GaN | InN |
|---|---|---|---|
| $a$ (Å) | 3.112 | 3.189 | 3.533 |
| $c$ (Å) | 4.98 | 5.186 | 5.693 |
| $E_g$(eV) | 6.2 | 3.42 | 1.89 |

To simulate the spin-glass state of $A_{1-x}V_xN$ (A= Al, Ga, In), the cations A are substituted randomly by $V^{\uparrow}_{x/2}$ and $V_{\downarrow x/2}$ magnetic ions, where ↑ and ↓ denote the direction of the local moment of V ions. Hence, the ferromagnetic and spinglass states can be described as $A_{1-x}V^{\uparrow}_xN$ and $A_{1-x}V^{\uparrow}_{x/2}V_{\downarrow x/2}N$, respectively. Then, the energy difference of the spinglass state and ferromagnetic state is estimated by $\Delta E = E_{sg}-E_{ferro}$, where $E_{sg}$ denotes the total energy of the spinglass state and $E_{ferro}$ that of the ferromagnetic state. The estimation of Curie temperatures is performed by using a mapping on the Heisenberg model in a mean field approximation [2]: $k_BT_c = 2\Delta E/3x$, where $k_B$ is Boltzmann constant, $x$ the concentration of substitutional Vanadium atoms.

The DOS for $Al_{1-x}V_xN$, $Ga_{1-x}V_xN$ and $In_{1-x}V_xN$ is calculated for several concentrations of Vanadium (2%, 5% and 10%) to explain the change of the exchange interaction mechanism corresponding to different concentrations of substituting magnetic ions. The Curie temperatures are evaluated through the calculation of the total energy differences $\Delta E$ for various concentrations of Vanadium (up to 18%).

## III. Density of states and Curie temperature

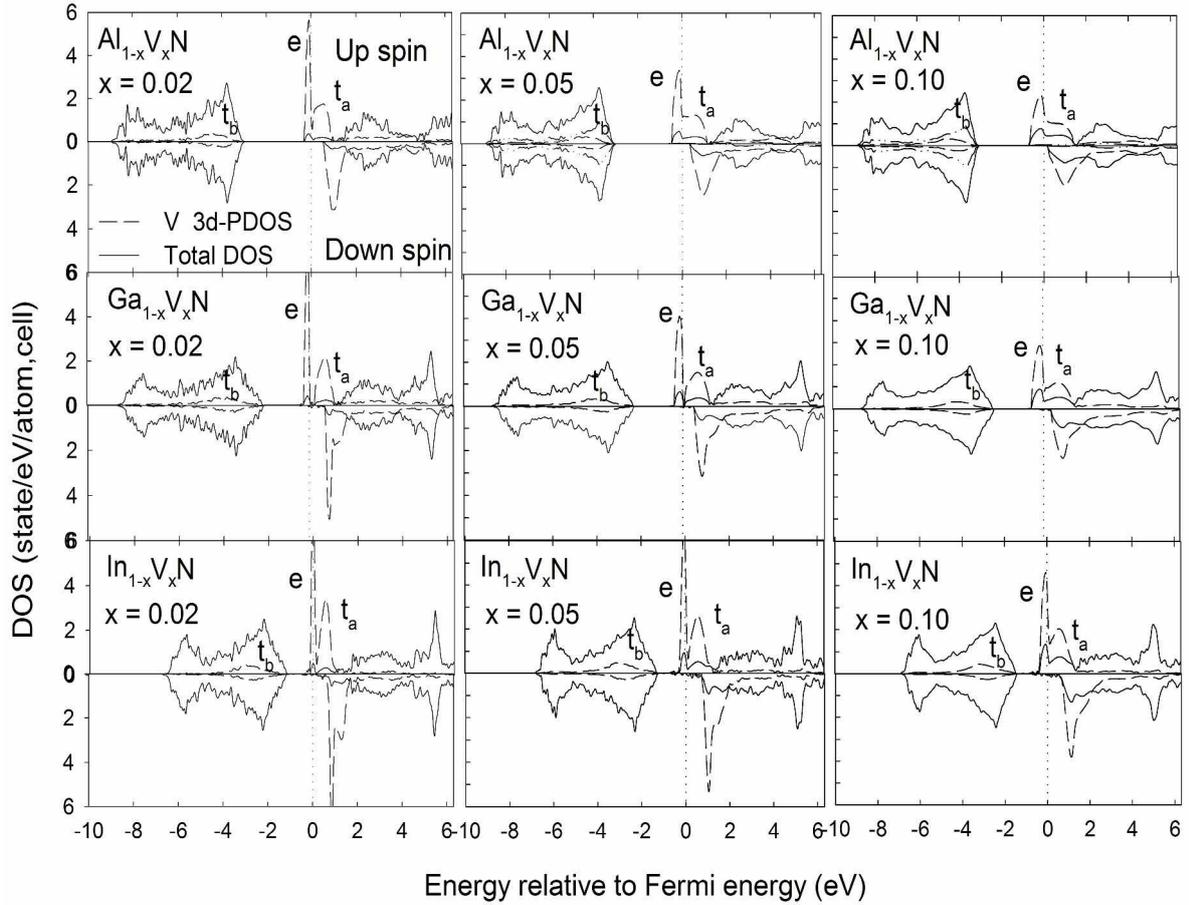

**Fig. 1.** Total DOS per unit cell (solid line), partial DOS of *3d*-states per V atom at cation site (dashed line) for $Al_{1-x}V_xN$, $Ga_{1-x}V_xN$ and $In_{1-x}V_xN$ at several V concentrations ($x = 0.02$, 0.05 and 0.10).

In order to investigate the systematic behavior of the magnetic states and mechanism of exchange interaction in $Al_{1-x}V_xN$, $Ga_{1-x}V_xN$ and $In_{1-x}V_xN$, the densities of states of considering DMS at several concentrations of V ($x = 0.02$, 0.05, and 0.10) are illustrated in Fig. 1. Energy is relative to Fermi energy. The partial DOS (PDOS) of *3d*-states of Vanadium was calculated per atom, and the total DOS was estimated per unit cell of the wurtzite structure that contains two molecular units of AN (A = Al, Ga, In). Valence bands of all compounds in question are dominated by *2p* states of Nitrogen. As shown in the figure, near the Fermi level, the impurity states formed by *3d*-electrons of Vanadium appear. With $d^2$ configuration of $V^{+3}$, the *3d*-states of Vanadium are split into three states: the bonding state ($t_b$) located in the host valence band, non-bonding (*e*) and anti-bonding state ($t_a$) in the midgap band. These impurity states show a large exchange splitting, leading to the high-spin configuration of *d* electrons. The *e* state of up spin locates just below the Fermi level as a sharp peak and fully occupied, while $t_a$ state locates right above Fermi level and slightly mixed with conduction band at high V concentrations. With increasing V concentration, two these states lightly mix to each other and create an impurity band at Fermi level. At high concentration of V, the anti-bonding state $t_a$ of V-*3d* electrons weakly hybridizes with *2p*-state of Nitride to make an impurity band. This band is broadened with increasing V concentrations.

At $x = 0.02$, the *e* and $t_a$ states separate from each other, leading to the *3d*-DOS at Fermi level $n_{3d}(E_F)$ becomes very low. With increasing $x$ (0.05, 0.10), $Al_{1-x}V_xN$ shows the strongly mixed between these states to create an impurity band at Fermi level. The mixing is weaker in $Ga_{1-x}V_xN$, and becomes weakest in $In_{1-x}V_xN$ due to the hybridization between V-*3d* and N-*2p* states decreases with increasing distant between atoms at lattice sites corresponding to the increase of lattice constant (see Tab. I). On the contrary, the splitting between *3d* states of up spin and down spin increases with lattice constant as shown in DOS of $Al_{1-x}V_xN$, $Ga_{1-x}V_xN$ and $In_{1-x}V_xN$.

Since the Fermi level lies between the $e$ and $t_a$ states, the density of states of *3d*-electrons at Fermi level $n_{3d}(E_F)$ is considered small at low concentrations of Vanadium, resulting in the weakly stable of the ferromagnetic states in the viewpoint of carrier induced ferromagnetic double exchange mechanism. However, the estimation of total energy differences and $T_c$ shows the stability of the ferromagnetic state and sufficient high $T_c$ even if Vanadium concentration is at several percent (Fig. 2). Hence, it differs from (Mn, Cr)-doped Nitrides, in which the double exchange mechanism is dominant [2, 16, 17], that the other mechanism called the ferromagnetic super-exchange is dominant one in V-doped Nitrides. Corresponding to this mechanism, while the $e$ state is fully occupied, the $t_a$ state is empty.

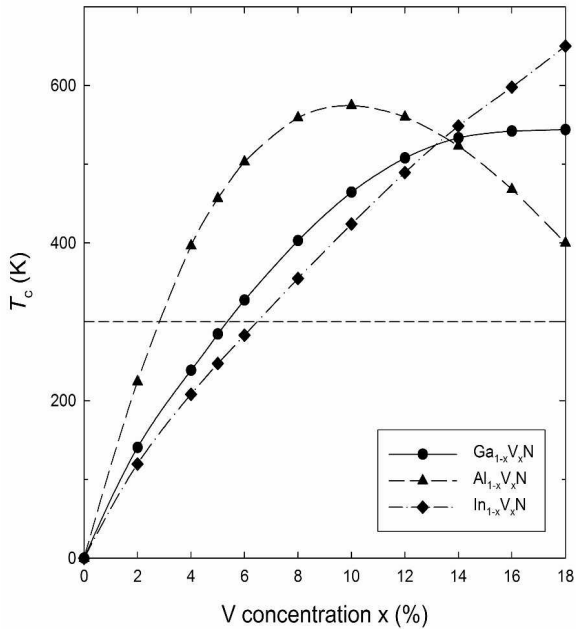

**Fig. 2.** $T_c$ *vs* Vanadium concentration for $Al_{1-x}V_xN$, $Ga_{1-x}V_xN$ and $In_{1-x}V_xN$.

Fig. 2 shows the Curie temperature estimated within MFA for $Al_{1-x}V_xN$ (dashed line), $Ga_{1-x}V_xN$ (solid line) and $In_{1-x}V_xN$ (dash-dotted line). In the DMS with the double-exchange mechanism being dominant, the Curie temperature is proportional to the broaden width of the peak of the anti-bonding states $t_a$ at Fermi level, and $T_c$ varies as a function of $x^{1/2}$. In the compounds in question, the $T_c$ has a nearly linear form of $x$ ($T_c \propto x$) at the dilute regime ($x < 0.08$ for $Al_{1-x}V_xN$, $x < 0.12$ for $Ga_{1-x}V_xN$) due to ferromagnetic super-exchange interaction. While the $T_c$ of $In_{1-x}V_xN$ increases monotonously for whole of concentrations, the $T_c$ of $Ga_{1-x}V_xN$ gains the saturation at $x = 0.14$ and that of $Al_{1-x}V_xN$ has a tendency of decrease with concentrations higher than 8%.

For the case of $Al_{1-x}V_xN$, the $T_c$ at $x > 0.08$ is nearly proportional to $- x$ corresponding to the dominant of the anti-ferromagnetic super-exchange interaction mechanism ($T_c \propto - x$). As seen from Fig. 1, because of the large lattice constant, the mixing between $e$ and $t_a$ states is quite small in $In_{1-x}V_xN$ and the ferromagnetic super-exchange interaction is more preferable than that in $Al_{1-x}V_xN$ and $Ga_{1-x}V_xN$. On the contrary, $Al_{1-x}V_xN$ has a higher $T_c$ (at $x < 0.14$) than $Ga_{1-x}V_xN$ and $In_{1-x}V_xN$ due to strong hybridization of N-*2p* states and V-*3d* states. However, because the lattice constants of AlN are smaller than that of GaN and InN, the anti-ferromagnetic super-exchange interaction effect in $Al_{1-x}V_xN$ is strongest and easier to suppress the ferromagnetic interaction, hence reduce the $T_c$. $Ga_{1-x}V_xN$ is the middle case, and the anti-ferromagnetic exchange interaction effect in $In_{1-x}V_xN$ is smallest. With $x > 0.04$, all of these compound are expected to have the $T_c$ exceeding the room temperature.

**IV. Conclusion**

In this paper, based on the *ab initio* calculations of electronic structures and Curie temperatures, a systematic investigation of magnetism of V-doped Nitrides ($Al_{1-x}V_xN$, $Ga_{1-x}V_xN$ and $In_{1-x}V_xN$) is presented. Electronic structures are calculated within KKA-LSDA-CPA method. The Curie temperature is estimated by using a mapping on the Heisenberg model in a mean field approximation. The obtained results show that the Curie temperature of V-doped Nitrides can be expected to exceed the room temperature with a few concentrations of Vanadium. In the dilute regime, the ferromagnetic super-exchange interaction mechanism is dominant for all considered compounds, and $T_c$ is nearly proportional to the substitutional concentration ($T_c \propto x$). With increasing $x$ higher than 0.10, the anti-ferromagnetic super-exchange interaction appears, suppresses the ferromagnetic interaction, resulting in the reduction of the $T_c$. Especially, the effect of this mechanism is strongest in $Al_{1-x}V_xN$ leading to $T_c$ decreases as a function of $- x$.

It is suggested that $T_c$ estimated within MFA could be higher than the room temperature at $x > 0.03$ for $Al_{1-x}V_xN$, and at $x > 0.05$ for $Ga_{1-x}V_xN$ and $In_{1-x}V_xN$, therefore we can expected a possibility of fabrication of V-doped Nitrides for spintronics device applications.

**Acknowledgments**

This research is supported by Special Fund for Co-ordination of Science and Technology and SANKEN-COE from the Ministry of Education, Cultural, Sport, Science and Technology. We thank to Prof. Akai H (Osaka University) for providing the useful KKR-CPA package.